\documentstyle[aps,prl,twocolumn,epsfig]{revtex}

\def\tm{\tablenotemark}

\begin{document}
\draft
\title{All-electron $GW$ approximation with the mixed basis
expansion based on the full-potential LMTO method}
\author{Takao Kotani$^1$ and Mark van Schilfgaarde$^2$}
\address{Department of Physics,
Osaka University, Toyonaka 560--0043, Japan$^1$}

\address{Sandia National Laboratory, Livermore, CA U.S.A.$^2$}
\date{\today}
\maketitle
\begin{abstract}
We present a new all-electron, augmented-wave implementation of
the $GW$ approximation using eigenfunctions generated by a recent
variant of the full-potential LMTO method.  The dynamically
screened Coulomb interaction $W$ is expanded in a mixed basis set
which consists of two contributions, local atom-centered
functions confined to muffin-tin spheres, and plane waves with
the overlap to the local functions projected out.  The former can
include any of the core states; thus the core and valence states
can be treated on an equal footing.  Systematic studies of
semiconductors and insulators show that the $GW$ fundamental
bandgaps consistently fall low in comparison to experiment, and
also the quasiparticle levels differ significantly from other,
approximate methods, in particular those that approximate the
core with a pseudopotential.
\end{abstract}


\narrowtext

\label{intro}


The $GW$ approximation (GWA) of Hedin\cite{Hedin1} has been applied to many
kinds of materials \cite{review_gw1,review_gw2}.  In customary {\it ab
initio} implementations, the self-energy $\Sigma$ is generated from
eigenvalues and eigenfunctions calculated within the self-consistent
local-density approximation (LDA).  It has been shown that quasiparticle
energies computed in this way are in significantly better agreement with
experiment that are the LDA eigenvalues.

The various implementations of GWA by may classified by what kinds of basis
sets are used in the expansion of the LDA eigenfunctions, and the expansion
of the bare and screened Coulomb interactions $v$ and $W$.  The most common
implementations make an additional pseudopotential approximation for the
core, which makes it possible to expand all these quantities in plane
waves.  However, a plane-wave basis is poorly suited to localized orbitals
such as $d$- or $f$- states. Moreover, as we show here, it appears that the
pseudopotential approximation is somewhat inadequate when used in
conjunction with the $GW$ approximation.  Two other $GW$ implementations
that do not use pseudopotentials have also been
published\cite{hamada90,Arnaud00}.  In both of these methods, $v$ and $W$
are expanded in plane waves; owing to the difficulty in a plane-wave
expansion of localized orbitals they did not take into account core
contributions. Aryasetiawan and collaborators implemented a method that
expands $v$ and $W$ in a linear combination of augmented wave function
products (product basis) and has applied it to several kind of materials,
including NiO\cite{aryasetiawan95}, with reasonable results.  However this implementation
requires the atomic spheres approximation (ASA) for the LDA, which
approximates space by a superposition of atom-centered (``muffin-tin'')
spheres, neglecting the interstitial.  Thus, its reliability is uncertain.

We present a new $GW$ implementation which uses a mixed-basis expansion for
$v$ and $W$.  $v$ and $W$ are expanded in Aryasetiawan's product-basis in
the muffin-tin (MT) spheres, and the interstitial plane waves (IPW) in the
interstitial region.  An IPW is a plane wave with the MT contributions
projected out.  This basis can be an efficient one applicable for the
localized electrons and cores.  Together with the rather accurate
eigenfunctions given by the all-electron full-potential LMTO method, our
$GW$ method can be both efficient and accurate.  It is our knowledge the
first implementation of $GW$ that makes no significant approximations
beyond the lack of self-consistency (pseudopotential approximation for the
core, shape approximation in the potential, plasmon-pole approximation for
the dynamical screening).

The $GW$ self-energy is
\begin{equation} \label{self_energy}
\Sigma({\bf r},{\bf r}^{\prime},\omega)=\frac{i}{2\pi}\int d\omega'
G({\bf r},{\bf r}^{\prime},\omega+\omega^{\prime})e^{i\delta\omega^{\prime}}
W({\bf r},{\bf r}^{\prime},\omega^{\prime}),
\end{equation}
where $W$ is the RPA screened Coulomb interaction and $G$ is the Green's
function.  $W$ and $G$ are calculated directly from the self-consistent
LDA eigenvalues and eigenfunctions, with exchange-correlation potential
$V_{xc}({\bf r})$.  Then the quasi-particle energy $E_n({\bf k})$ ($n$ is the
band index and ${\bf k}$ is the wave vector) is given by
\begin{eqnarray}
\label{quasiparticle_energy}
&&E_n({\bf k})-\epsilon_n({\bf k})=Z_{n{\bf k}} \times \nonumber \\
&&[\langle\Psi_{{\bf k}n}|
\Sigma({\bf r},{\bf r}^{\prime},\epsilon_n({\bf k}))|\Psi_{{\bf k}n}\rangle
- \langle\Psi_{{\bf k}n}|V_{\rm xc}^{\rm LDA}(r)|\Psi_{{\bf k}n}\rangle]
\end{eqnarray}
where $\epsilon_n({\bf k})$ and $\Psi_{{\bf k}n}$ denotes the LDA
eigenvalues and eigenfunctions,
and $Z_{n{\bf k}}$ is the quasi-particle (QP) renormalization factor
\begin{equation}\label{Renormalization}
Z_{n{\bf k}}=[1-\langle\Psi_{{\bf k}n}|
\frac{\partial}{\partial\omega} \Sigma({\bf r},{\bf r}^{\prime},
\epsilon_n({\bf k}))|\Psi_{{\bf k}n}\rangle]^{-1}.
\end{equation}

Our method to calculate the quasi-particle energy, is based on the
FP-LMTO method which expands the eigenfunction $\Psi_{{\bf k}n}$ (we
omit the spin index to simplify expressions) as
\begin{equation}\label{eigenfunction}
\Psi_{{\bf k}n} = \sum u^{{\bf k}n}_s \chi^{{\bf k}}_s({\bf r}).
\end{equation}
Here $\chi^{{\bf k}}_s({\bf r})$ denotes the Bloch sum of the MT orbitals
$\chi_s({\bf r})$.  $s \equiv (a,n,L)$ is a composite index labelling the
atom in the unit cell $a$, the angular momentum $L$, and another index $n$
to specify which MT orbitals (multiple envelope functions per $L$ are
permitted) In the interstitial, $\chi^{{\bf k}}_s({\bf r})$ may be expanded
in plane waves, while at the augmentation spheres it is matched smoothly
and onto of a linear combination $A_{au}({\bf r}) \equiv \{ \phi_{aL}(r)
Y_L(\hat{\bf r}),\dot{\phi}_{aL}(r) Y_L(\hat{\bf r}) \}$ of solutions of
the radial Schr\"odinger equation, $\phi_{aL}(r)$, and its energy
derivative, $\dot{\phi}_{aL}(r)$.  Finally, $u \equiv(L, I_{\rm P})$ is a
composite of $L$ and $I_{\rm P}$, where $I_{\rm P}$ assumes value 0 for $\phi$
and 1 for $\dot{\phi}$.  Therefore $\Psi_{{\bf k}n}$ can be written as
\begin{eqnarray}
&& \Psi_{{\bf k}n} ({\bf r})
 = \sum_{a u} \alpha^{{\bf k} n}_{au} A^{\bf k}_{a u}({\bf r})
 + \sum_{\bf G} \beta^{{\bf k}n}_{\bf G} P^{\bf k}_{\bf G}({\bf r}) \label{psieq},
\end{eqnarray}
where the IPW $P^{\bf k}_{\bf G}({\bf r})$ is defined as
\begin{eqnarray}
P^{\bf k}_{\bf G}({\bf r}) &=& 0  \ \ \ {\rm \ if \ {\bf r} \in Any \ MT} \nonumber \\
        &=&   \exp (i ({\bf k+G}) {\bf r}) \ \ \ {\rm Otherwise}.
\end{eqnarray}
In the interstitial, it is evident that products $\Psi_{{\bf k_1}n} \times
\Psi_{{\bf k_2}n'}$ can similarly be expanded in IPW's because $P^{\bf
k_1}_{\bf G_1}({\bf r}) \times P^{\bf k_1}_{\bf G_2}({\bf r})= P^{\bf
k_1+k_2}_{\bf G_1+G_2}({\bf r})$.  Within the augmentation sphere products
of $\Psi$ are expanded in $B_{am}^{\bf k_1+k_2}({\bf r})$, which is the
Bloch sum of the product basis $B_{am}({\bf r})$.  They are constructed
from the products of $A_{au}({\bf r}) \times A_{a'u'}({\bf r})$ following
the procedure by Aryasetiawan\cite{aryasetiawan94}.

Thus a basis suitable for expansion of wave function products, and
therefore also $v$ and $W$, is the mixed basis $\{M^{\bf k}_I({\bf r})
\}\equiv \{ \tilde{P}^{\bf k}_{\bf G}({\bf r}), B_{am}^{\bf k}({\bf r})\}$,
where $\tilde{P}^{\bf k}_{\bf G}$ is an orthonormal basis rendered
from $P^{\bf k}_{\bf G}$. $I\equiv \{ {\bf G},am\}$ is a composite index
labelling the basis.  The complete information to calculate the self-energy
and $E_n({\bf k})$ are the Coulomb matrix $v_{IJ}({\bf k})= \langle M^{\bf
k}_I |v | M^{\bf k}_J \rangle$, the matrix elements of the products
$\langle \Psi_{{\bf q}n}| \Psi_{{\bf q-k}n'} M^{\bf k}_I \rangle$ and the
eigenvalues $\epsilon_{{\bf k}n}$.  The exchange part of the self-energy is
written as
\begin{eqnarray}
\langle \Psi_{{\bf q}n}|\Sigma_{\rm x} |\Psi_{{\bf q}n} \rangle
&=&\sum^{\rm BZ}_{{\bf k}}  \sum^{\rm  occ}_{n'}
\langle \Psi_{{\bf q}n}| \Psi_{{\bf q-k}n'} M^{\bf k}_I \rangle v_{IJ}({\bf k}) \times \nonumber \\
&&\langle M^{\bf k}_J \Psi_{{\bf q-k}n'} | \Psi_{{\bf q}n} \rangle.
\end{eqnarray}
The screened Coulomb interaction $W_{IJ}({\bf q},\omega)$ is calculated
through $W = (1-v D)^{-1} v$, where the polarization function $D$ is
\begin{eqnarray}
&&D_{IJ}({\bf q},\omega)
=\sum^{\rm BZ}_{{\bf k}}  \sum^{\rm  occ}_{n} \sum^{\rm  unocc}_{n'}
\langle M^{\bf k}_I \Psi_{{\bf q}n} |\Psi_{{\bf q-k}n'} \rangle 
\langle \Psi_{{\bf q}n}| \Psi_{{\bf q-k}n'} M^{\bf k}_J \rangle \nonumber \\
&&\times
(\frac{1}{\omega-\epsilon_{{\bf k}n}+\epsilon_{{\bf q-k}n'}+i \delta}
-\frac{1}{\omega+\epsilon_{{\bf k}n}-\epsilon_{{\bf q-k}n'}-i \delta}). \label{dieele}
\end{eqnarray}
Finally, the correlation part of self-energy is calculated by
\begin{eqnarray}
&&\langle \Psi_{{\bf q}n}|\Sigma_{\rm c}(\omega) |\Psi_{{\bf q}n} \rangle
= \sum^{\rm BZ}_{\bf k}  \sum^{\rm  All}_{n'} \sum_{IJ}
\langle \Psi_{{\bf q}n}| \Psi_{{\bf q-k}n'} M^{\bf k}_I \rangle \nonumber \\
&&\ \  \times \langle M^{\bf k}_J \Psi_{{\bf q-k}n'} | \Psi_{{\bf q}n} \rangle  \nonumber \\
&&\ \  \times \int_{-\infty}^{\infty} \frac{i d\omega'}{2 \pi}
W_{IJ}({\bf k},\omega')
\frac{1}{\omega'+\omega-\epsilon_{{\bf q-k}n'} \pm i \delta}. \label{sc}
\end{eqnarray}
Here the denominator $-i \delta$ is for occupied states, and $+i \delta$ for unoccupied states.
We use the $\omega'$-integral method given by Aryasetiawan\cite{aryasetiawanasa1}.
Our $GW$ code is developed starting from his code.



Details of the method will be described elsewhere.  Applying the method to
a wide range of semiconductors, we find a number of systematic tendencies.
The most important ones are: $(i)$ there is a systematic underestimate of
the fundamental bandgap $(ii)$ The bandgap error tends to increase with
bond polarity, and for compounds with relatively shallow cation $d$ states.
$(iii)$ the GW rather badly underestimates the deepening (relative to LDA)
of occupied cation $d$ states found in III-V and II-V compounds.  $(iv)$
For zincblende lattices, the bandgap error tends to be slightly larger at
the X point than at $\Gamma$. $(v)$ the bandgap is sensitive to what
approximations are used for shallow core states.

GaAs illustrates many of these general findings.  It has a direct gap at 0K
of 1.52~eV; however to compare against the present non spin-polarized
calculations, we assign the valence band maximum $\Gamma_{15}$ to
$\frac{1}{3}\Gamma_{7v}+ \frac{2}{3}\Gamma_{8v}$.  (Spin-orbit coupling
splits $\Gamma_{15}$ by shifting the $\Gamma_{8v}$ states
$+\frac{1}{3}\Delta_0$ and the $\Gamma_{7v}$ state $-\frac{2}{3}\Delta_0$.
This was confirmed by an LDA-ASA calculation including spin-orbit coupling,
which also yielded $\Delta_0$=0.36~eV, in good agreement with the observed
$\Delta_0$=0.34~eV.)  Table \ref{table:gaaslevels} shows gaps at $\Gamma$,
X and L; the spin-orbit-corrected fundamental gap is 1.63~eV.  An accurate
LDA calculation must include both 3$d$ states (which push upwards on
$\Gamma_{15}$, narrowing the gap) and the 4$d$ states (which push
downwards, widening the gap).  At present the $GW$ implementation can
include only one of these states for the input wave functions (the $GW$
itself includes both), but to compute the bands within our present $GW$ we
must select either 3$d$ or 4$d$ to input to the $GW$.  Therefore, all three
LDA cases are presented (the 3+4$d$ case uses local
orbitals\cite{singh91}), and are seen to fall within $\sim$0.15~eV of each
other.

Similarly the $GW$ gaps, whether the LDA 3$d$ basis or 4$d$ basis is input,
lie within $\sim$0.15~eV of each other (albeit with a slightly different
$k$-dependence).  By comparing LDA bands to the 3+4$d$ case, the 4$d$ is a
better choice for GaAs, especially in the conduction bands where the Ga
4$d$ begin to play an important role.  In any case, the direct gap
($\Gamma_{15v}\rightarrow\Gamma_{1c}$) is $\sim$0.3~eV smaller than the
experimental value; while the X and L point (X$_{1c}$ and X$_{1c}$) are
underestimated by a somewhat larger amount, $\sim$0.4~eV.  We have found
this tendency to be rather systematically followed in the III-V
semiconductors, as shown below.

The effect of the core is particularly important.  Line {\it core}0 in
Table \ref{table:gaaslevels} shows what gaps result when valence states
only are included in the calculation of $D$ and $\Sigma$.  For this case
the LDA potential we subtract corresponds to the valence-only density,
$V^{\rm LDA}_{\rm xc}(n_{\rm val};{\bf r})$.  This approximates what is
typically done in pseudopotential $GW$ (GW:PP) calculations.  It is seen
that this approximation leads to much too strong a $k$-dependence on the
gap shift, something also seen when compared to GW:PP calculations.  Table
\ref{table:gaaslevels} shows some data of Shirley et. al.\cite{shirley97},
(marked GW:PP).  They rather closely track the {\it core}0 results except
for an approximately $k$-independent shift of 0.32~eV.  Shirley also
included an approximate core polarization term (marked GW:PP+CP).  It is
seen that the addition of core-polarization terms to the PP have roughly
similar effects (increasing the shift at $\Gamma$, decreasing it at X), but
the shifts are larger than the all-electron results, and depend more
strongly on $k$.  Similar overestimates of gaps by the GW:PP method are
seen in Si and AlAs.  For example Shirley computed the
$\Gamma_{15v}${}$\rightarrow$X$_{1c}$ transition to be 1.31~eV, while our
all-electron result is 1.04~eV, close to the GW:PAW result (1.10~eV) of
Arnaud\cite{Arnaud00}, and an early GW:LAPW result (1.14~eV) by
Hamada\cite{hamada90}.  Moreover, we find for GaN (and indeed generally for
all semiconductors studied), a weak energy-dependence of the $GW$-LDA shift
of the conduction bands up to 10 eV above the conduction-band minimum, in
accord with a careful analysis of both UV reflectivity and near-edge x-ray
absorption spectra\cite{lambrecht97}.  However, using the GW:PP method,
Rubio et. al\cite{rubio93} found a significant energy-dependence of this
shift.

Inspecting the bottom of Table \ref{table:gaaslevels}, it is evident that
the Ga 3$d$ states contribute in an important way to the dielectric
response.  The row marked {\it core}2 shows that the levels change rather
significantly when the Ga 3$d$ is omitted from the calculation of $D$.

\begin{figure}
\begin{center}
\mbox{\epsfig{file=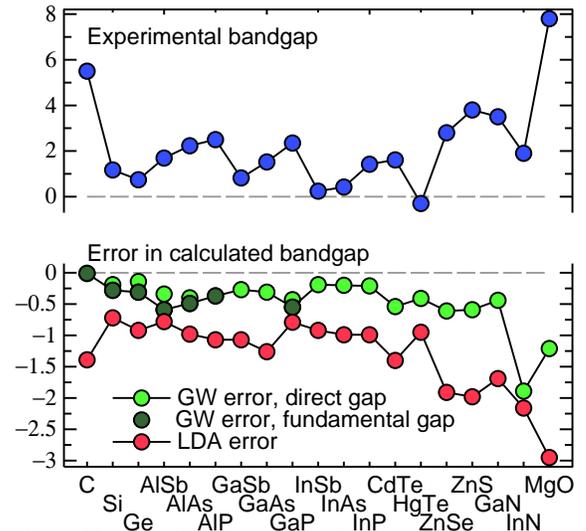,height=7cm}}
\caption{Top panel: experimental values of the fundamental gap in a
variety of semiconductors and insulators, approximately ordered by ionic
character. Bottom panel: deviations from experiment in the fundamental
gap. Red points are LDA errors; green points are $GW$ errors.  The light
and green denotes the smallest direct gap; the dark green denotes
the fundamental gap when it differs from the direct gap.
A term $1/3\Delta_0$ was subtracted from the calculated $GW$ and
LDA data to account for spin-orbit coupling, as discussed in the text.
\label{fig:gaps}}
\end{center}
\end{figure}

Fig.~\ref{fig:gaps} illustrates the systematics of the errors in the
fundamental gap for a range of semiconductors and insulators.  With the
sole exception of diamond\cite{noteppgwdiamond}, the $GW$ gaps are
systematically smaller than experiment.  As noted for GaAs above, there is
some $k$-dependence to the errors: the lowest conduction band levels at L
and more especially at X are generally in worse agreement than those at
$\Gamma$.  Two sources of error are readily identifiable, and as will be
shown elsewhere\cite{marksx}, they account for nearly all of the gap errors
in these $sp$ bonded semiconductors.  
As noted by Maksimov\cite{Maksimov},
an important contribution to the gap error arises from the nonlocal
screened exchange missing from LDA.  In Hartree-Fock theory, the bare
exchange is used, therefore it rather severely overestimates the bandgap and
bandwidths in general.  The $GW$ does screen the exchange, and
qualitatively speaking $GW$ replaces in addition to $V_{\rm xc}^{\rm LDA}$,
a term which corresponds to the nonlocal screened exchange, 
with the nonlocal screened exchange written as
\begin{equation}\label{epsiloninterpolation}
\Sigma \approx V_{\rm xc}^{\rm LDA} + \frac{1}{\epsilon_{\infty}} (\Sigma_{\rm x} - V_{\rm x }^{\rm  LDA}).
\end{equation}

In the present non self-consistent $GW$ implementation, $\epsilon$ is
computed from the LDA, with $\epsilon_{\infty}$ $\sim$10 to 20\%
overestimated on account of the LDA gaps being too small.  As we will show
elsewhere\cite{marksx}, self-consistency reduces $\epsilon$ slightly.  This
enhances the nonlocal part of the self-energy operator, and further
increases the $GW$-LDA gap correction, as can be qualtitatively understood
by Eq.~\ref{epsiloninterpolation}.  This correction is 0.2~eV in
Si\cite{weiku} but increases in the more ionic materials where $\epsilon$
is smaller.  That the correction is larger in the latter case might be
expected because the initial LDA potential (from which $GW$ is constructed)
is a poorer approximation, as again can be qualitatively seen from
Eq.~\ref{epsiloninterpolation}.

Table \ref{table:cationdlevels} shows the underbinding by $GW$ of the
occupied cation $d$ levels.  As is shown, the LDA rather badly
underestimates this binding.  The LDA error is well known and also
reasonably well understood\cite{lambrecht94}, namely that the LDA
eigenvalues are inappropriately interpreted as excitation energies.  It is
significant that, while the $GW$ shifts these levels in the direction of
experiment, the shift is much too small.  Comparison with the exchange-only
(Hartree-Fock) column shows clearly that the $GW$ is strongly overscreening
the bare exchange for these cation $d$ levels.  (Similar difficulties are
found for the so-called charge transfer antiferromagnets such as NiO and
MnO, as will be discussed elsewhere.)  This underbinding also contributes
an important term to the gap error; shows importance increases for shallow
$d$ states.  It is not clear whether self-consistency in the $GW$ will
remedy this error.  It is interesting that, even within the LDA, these
levels can be reasonably computed using Slater transition-state theory.

The underbinding of the $d$ levels makes an important contribution to the
gap error where the levels are relatively shallow (e.g. CdTe, and GaAs).
There is a coupling between the $d$ state and and the valence band maximum,
which pushes the latter upwards and reduces the bandgap.  When the energy
separation between these states are too small, the coupling them is
overestimated.  This will be shown in some detail in a future
work\cite{marksx}. In additon, we will publish the GW results 
on the wurzite ZnO where we prepare the input eigenfunctions
by the full-potential LAPW method \cite{usuda}.

In summary, an all-electron implementation of $GW$ presented here leads to
systematic errors in semiconductor and insulator bandgaps, whose origin
were identified.

\vbox{
\begin{table}
\caption{Selected energy eigenvalues, in eV, at $\Gamma$, L and X
for GaAs.  Experimental data are taken from
Ref.~\protect\cite{landolt}; the line ``Expt-SO'' subtracts the
spin-orbit coupling from the raw experimental data to compare
with the calculated levels, as discussed in the text.  The
GW:PP+CP and GW:PP data are GW results\protect\cite{shirley97}
using an LDA pseudopotential, with and without a core
polarization term added.  The GW results are quoted with the $Z$
(underlined), and with $Z$=1 immediately below.  Also shown are
changes in the GW levels owing to varying degrees of
approximation for the Ga and As $3d$ states.
\label{table:gaaslevels}}
\begin{tabular}{ l|r|r|r|r|r}
\vbox to 12pt{}
          & $\Gamma_{1c}$ & $\Gamma_{15c}$ & L$_{1c}$ & X$_{1c}$ & X$_{3c}$ \\
\tableline
\tableline
Expt      & 1.52  & 4.72 &  1.84  & 2.01  & 2.41\vbox to 12pt{}\\
Expt-SO   & 1.63  & 4.83 &  1.95  & 2.12  & 2.52\\
LDA       & 0.35  & 3.68 &  0.86  & 1.34  & 1.54\vbox to 12pt{}\\
LDA(4d)   & 0.41  & 3.72 &  0.91  & 1.34  & 1.58\\
LDA(3d)   & 0.24  & 3.57 &  0.76  & 1.33  & 1.57\\
GW,4d(Z)  & 1.35  & 4.20 &  1.54  & 1.56  & 1.97\vbox to 12pt{}\\
\tableline
GW,4d(Z=1)& 1.50  & 4.28 &  1.63  & 1.59  & 2.02\\
GW,3d(Z)  & 1.34  & 4.38 &  1.59  & 1.71  & 2.04\\
GW,{\it core}0\tm[4]
          & 0.94  & 4.18 &  1.37  & 1.72  & 2.01\\
GW:PP     & 1.29  &      &  1.69  & 2.05  & 2.34\vbox to 12pt{}\\
GW:PP+CP  & 1.69  &      &  1.88  & 1.93  & 2.29\\

HF,4d     & 4.84  & 8.50 &  5.10  & 5.26  & 5.63\vbox to 12pt{}\\

$\delta$GW,{\it core}3\tm[1]
          & 0.02  & 0.03 &  0.02  & 0.03  & 0.04\vbox to 12pt{}\\
$\delta$GW,{\it core}2\tm[2]
          & 0.24  & 0.34 &  0.26  & 0.25  & 0.24\\
$\delta$GW,{\it core}1\tm[3]
          & 0.26  & 0.35 &  0.27  & 0.25  & 0.25\\
$\delta$GW,{\it core}0\tm[4]
          &-0.41  &-0.02 & -0.17  & 0.16  & 0.04\\

\end{tabular}
\tablenotetext[1]{{\it core}3 treats the As 3$d$ at the Hartree-Fock
level (bare exchange) and omits this state in the computation of $D$}
\tablenotetext[2]{{\it core}2 additionally omits the Ga 3$d$ in the
calculation of $D$}
\tablenotetext[3]{{\it core}1 additionally approximates the Ga 3$d$
at the Hartree-Fock level}
\tablenotetext[4]{{\it core}0 computes both
$D$ and $W$ with valence states only, completely neglecting
the core beyond the LDA treatment.}

\end{table}
}

\vbox{
\begin{table}
\caption{Occupied cation $d$ band energy levels in selected II-VI and III-V
semiconductors, relative to the valence band maximum.  Column `HF' is the
exchange-only (Hartree-Fock) result.
\label{table:cationdlevels}}
\begin{tabular}{ l|r|r|r|r}
\vbox to 12pt{}
          &  Expt        &  LDA  & $GW$  & HF \\
\tableline
GaN       & -17.1\tm[1]  &-13.6  & -16.4 & -23.3 \\
GaAs      & -18.8\tm[3]  &-15.0  & -18.1 & -27.2 \\
InP       & -16.8\tm[3]  &-14.2  & -15.7 & -21.2 \\
InAs      & -17.1\tm[3]  &-14.4  & -16.1 & -22.6 \\
ZnS       &  -8.7\tm[2]  & -6.2  &  -7.1 & -14.1 \\
ZnSe      &  -9.0\tm[3]  & -6.7  &  -7.7 & -15.2 \\
CdS(ZB)   &  -9.2\tm[4]  & -7.5  &  -8.2 & -12.3 \\
CdTe      & -10.5\tm[3]  & -8.1  &  -9.3 & -14.4 \\
HgTe      &  -8.6\tm[3]  & -7.1  &  -7.6 & -12.8 \\
\end{tabular}
\tablenotetext[1]{\onlinecite{lambrecht94}}
\tablenotetext[2]{\onlinecite{zhou97}}
\tablenotetext[3]{\onlinecite{ley74}}
\tablenotetext[4]{\onlinecite{stampfl97}, average
of X and $\Gamma$, ``turning points'' method.}
\end{table}
}

%
%
%
%

\begin{references}
\bibitem{noteppgwdiamond}
Diamond appears to be the one material where the PP:GW and all-electron
results agree.  This is not suprising, since diamond has only one (very
deep) core state.
\end{references}
\end{document}